\documentclass{aa}
\usepackage{graphics}
\usepackage{epsf}
\newcommand{\kms}{km\,s$^{-1}$}

\begin{document}
%\thesaurus{12(02.08.1; 02.19.1; 09.10.1; 09.13.2)}
\title{The mass-velocity and intensity-velocity relations in jet-driven
molecular outflows}
\author{Turlough P.\ Downes \inst{1}
\and Sylvie Cabrit \inst{2}
% \and Alejandro Raga \inst{3}
}
\titlerunning{Jet-driven molecular outflows}
\authorrunning{T.P.\ Downes \& S.\ Cabrit}

\offprints{T.\ P.\ Downes, turlough.downes@dcu.ie}
\mail{School of Mathematical Sciences, Dublin City University,
Glasnevin, Dublin 9, Ireland}
\institute{School of Mathematical Sciences, Dublin City University,
Dublin 9, Ireland
\and LERMA, Observatoire de Paris, 61 Av.\ de l'Observatoire, F-75014
Paris
% \and Instituto de Ciencias Nucleares, UNAM, Mexico
}
\date{Received June 18, 2001;accepted March 12, 2003}

\abstract{ We use numerical simulations to examine the mass-velocity and
intensity-velocity relations in the CO J=2-1 and H$_2$ S(1)1-0 lines for
jet-driven molecular outflows.  Contrary to previous expectations, we find
that the mass-velocity relation for the swept-up gas is a single
power-law, with a shallow slope $\simeq -1.5$ and no break to a steeper
slope at high velocities.  An analytic bowshock model with no post-shock
mixing is shown to reproduce this behaviour very well.

We show that molecular dissociation and the temperature dependence of
the line emissivity are both critical in defining the shape of the line
profiles at velocities above $\sim$ 20 km s$^{-1}$.  In particular, the 
simulated CO J=2-1 intensity-velocity relation does show a break in slope, even
though the underlying mass distribution does not.  These predicted CO
profiles are found to compare remarkably well with observations of
molecular outflows, both in terms of the slopes at low and high
velocities and in terms of the range of break velocities at which the
change in slope occurs. Shallower slopes are predicted at high velocity
in higher excitation lines, such as H$_2$ S(1)1-0.

This work indicates that, in jet-driven outflows, the CO J=2-1 intensity
profile reflects the slope of the underlying mass-velocity distribution
only at velocities $\le $ 20 \kms, and that higher temperature tracers
are required to probe the mass distribution at higher speed.

}
\maketitle

\section{Introduction}
\label{introduction}
Various authors have noted that the intensity-velocity relationship
observed in low-J CO lines in molecular outflows tends to follow a
broken power-law $I_{\rm CO}(v) \propto v^{-\gamma}$, with $\gamma
\approx$ 1.8 $\pm$ 0.5 up to line-of-sight velocities $v_{\rm break}
\approx$ 10-30 km\,s$^{-1}$ and $\gamma
\approx$ 3-7 at higher velocities (e.g. Rodr\'{\i}guez et al.
\cite{rodriguez}; Stahler \cite{stahler}; Bachiller \& Tafalla \cite{bataf}; 
Richer et al. \cite{richer}). This property is an important test for 
proposed mechanisms of molecular outflow acceleration. In particular, 
recent work has addressed this issue in the case of entrainment by a jet.

Using an analytic model of a jet/bowshock system, Zhang \& Zheng
(\cite{zz97}) predicted a mass distribution, $m(v)$, following a broken
power-law $v^{-\mu}$ with slopes $\mu \approx 1.8$ up to 10
km\,s$^{-1}$ and $\mu \approx 5.6$ beyond. The observed values of
$\gamma$ in CO intensity profiles are then reproduced, provided the CO
abundance and line excitation do not vary much with velocity. However,
this seems highly unlikely, given the broad range of shock strengths
in a bowshock.

Smith, Suttner \& Yorke (\cite{smith}) conducted numerical simulations of
jet-driven molecular outflows that take into account dissociation and
heating in shocks.  They find that $I_{\rm CO}(v)$ follows a power-law
$v^{-\gamma}$ with $\gamma\approx$1.2--1.6 up to 10 km\,s$^{-1}$, and
a steeper slope further out, in qualitative agreement with
observations. They attribute this result to a much steeper mass-velocity
relation, $m(v) \propto v^{-3.5}$, than Zhang \& Zheng (\cite{zz97}).  
However, their reasoning involves an erroneous high-temperature dependence 
of the CO emissivity ($\propto$ T instead of $T^{-1}$). Therefore, the actual
origin of the slope of $I_{\rm CO}(v)$ in jet simulations, and the
underlying mass-velocity relation itself, remain to be clearly 
established. In this work we clarify this issue using simulations at 
higher resolution, and analytical modelling.

\section{Numerical method}
\label{numerical_method}

The code used in this work is very similar to that used in Downes \& Ray 
(\cite{dr2}).  The initial conditions are also very similar, but we give
a brief overview of them here for completeness.

The simulations are performed in 2D cylindrical symmetry.  The
densities of molecular and atomic hydrogen are tracked, along with the
ionisation fraction of hydrogen.  The CO density is assumed to be a
constant ($10^{-4}$) fraction of the H$_2$ density by number.  The
numerical scheme is a Godunov scheme which is second order in time and
space (see, e.g., Downes \& Ray \cite{dr2}).

The grid-spacing was set at $10^{14}$ cm.  The jet radius was $5
\times 10^{15}$ cm, and its density was 100 cm$^{-3}$ (equal to the
ambient density), thus allowing a resolution of cooling layers a 
factor of $10^2$-$10^4$ higher than in Smith et al. (\cite{smith}).
The time-averaged jet velocity, $v_0$, was 215 km s$^{-1}$.  Superimposed 
on this were sinusoidal variations with periods of 5, 10, 20 and 50 
years with total amplitude $v_1$.

In order to explore the effects of varying the jet velocity, three
simulations were run, as follows: (1) a ``steady'' jet where ${v_1
\over v_0}=0$; (2) a ``pulsed'' jet where ${v_1 \over v_0} = 0.6$; and
(3) a ``ramped-up'' jet where ${v_1 \over v_0} = 0.6$ and $v_0$
increases linearly in time from 0 to 215 km s$^{-1}$ over 100 years.
Typically, simulations of jets start off with the jet propagating at
full speed into the ambient medium.  It is likely that YSO jets do not
`switch on' impulsively, and it has been suggested recently (Lim et
al.\ \cite{limetal2002}) that a slow start-up of a jet may lead to
increased molecular abundance at the head of the bowshock, which would then 
affect the resulting intensity-velocity relation that we wish to model.  

\section{Numerical results}
\label{numerical_results}

\begin{figure*}
\resizebox{\hsize}{!}{\includegraphics{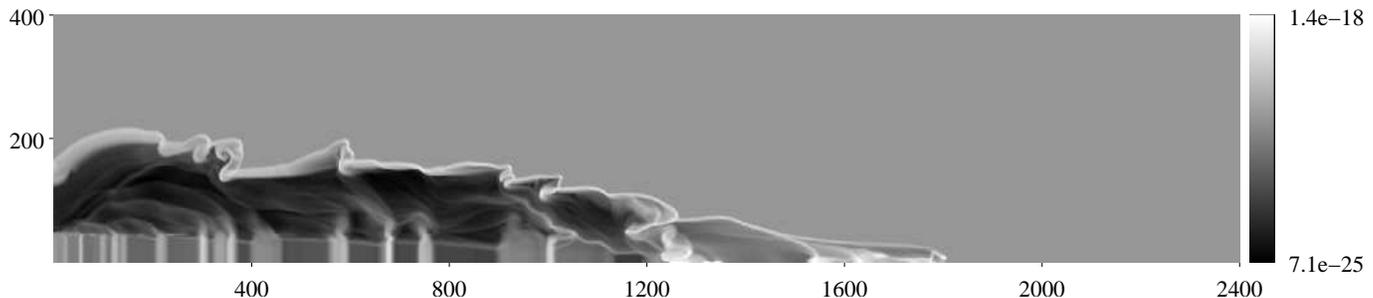}}
\caption{Log-scale plot of the distribution of number density for the
 pulsed jet with $\frac{v_1}{v_0} = 0.6$ at $t = 400$ yrs. The density scale 
is in units of g cm$^{-3}$ while the distance scales are in units of 
$1\times10^{14}$ cm
\label{den_f1}
}
\end{figure*}

In this section, we concentrate on the results of simulation (2)
(pulsed jet) described in Sect.\ \ref{numerical_method}, at a time
$t=400$ yrs. Results for other times and for the steady and ramped-up
jets (simulations (1) and (3) in Sect.\ \ref{numerical_method}) will
be discussed in Sect.\ \ref{effects}.

\begin{figure}
\resizebox{\hsize}{!}{\includegraphics{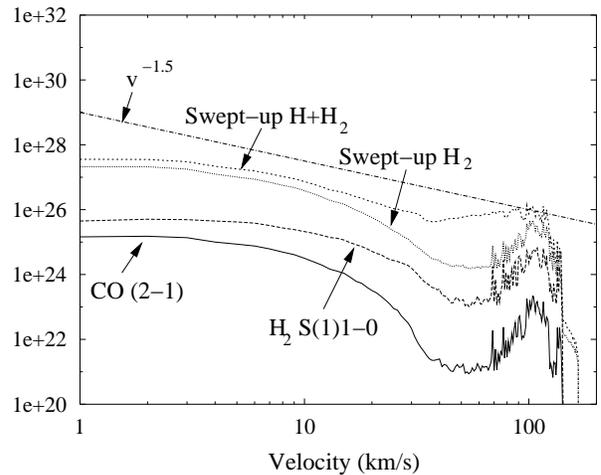}}
\caption{Plots of the mass-velocity relations for all swept-up material
(i.e.\ excluding jet material), $m(v)$, and for molecular swept-up material 
only, $m_{\rm H_2}(v)$, for the pulsed jet at $t=400$ yrs.  Also
shown are the intensity-velocity relations for the CO J=2-1 line and for
the H$_2$ S(1)1-0 line (arbitrary vertical offsets).  An angle of $30^\circ$ to
the plane of the sky is assumed.  \label{mv_iv}}
\end{figure}

Figure \ref{den_f1} contains a log-scale plot of the density
distribution for the chosen simulation. We can see that the
bowshock is rather irregular.  These irregularities are attributed to
the growth of the Vishniac instability (Downes \& Ray \cite{dr2}). In 
addition, several ``mini-bowshocks'' are present along the jet length, 
tracing internal working surfaces resulting from the variability of the jet 
velocity. Figure \ref{mv_iv} contains plots of various mass-velocity and
intensity-velocity relations for the same simulation, assuming an
inclination of $30^\circ$ to the plane of the sky.
We discuss each of these distributions in detail in the following.

\subsection{The total mass-velocity relation}
\label{num_totalmv}

A significant feature of Fig.\ \ref{mv_iv} is that $m(v)$, the
mass-velocity relation for all swept-up material (i.e.\ excluding jet
material), remains quite shallow across the whole velocity range. It
is essentially flat at the lowest velocities ($v < 3$ \kms), then
follows an approximate power-law $m(v) \propto v^{-\mu}$ with exponent
$\mu \simeq 1.5$ at intermediate velocities, before rising slightly
again above $v \simeq 40$ \kms.  The power-law slope that we find
agrees quite well with the value of $\mu \simeq 1.8$ predicted by the
analytical model of Zhang \& Zheng (\cite{zz97}) for low velocities,
but their predicted break to a steeper slope $\mu \simeq 5.6$ at
velocities above 10 \kms\ is not seen.

\subsection{The molecular mass-velocity relation}
\label{num_h2_mv}

The mass-velocity relation for swept-up {\it molecular} material,
$m_{\rm H_2}(v)$, behaves similarly to the total mass-velocity
relation, $m(v)$, at low velocities.  It does, however, become
considerably steeper than $m(v)$ at higher velocities (v $\ge$ 20
\kms).  This steepening occurs because material at these velocities
has been accelerated by faster shocks near the apex of the bowshock,
where significant molecular dissociation occurs. Indeed, the typical
dissociation limit for low-density non-magnetic shocks is $\simeq$ 30
\kms\ (see e.g. Flower et al. 2003). We note that the molecular
fraction appears to rise again at the highest velocities (above 80
\kms). This is an unavoidable result of numerical diffusion at the bow head,
which results in mixing of molecular jet material into the high-velocity
swept-up gas.

\subsection{The $I_{\rm CO}(v)$ relation}
\label{num_cov_relation}

The CO J=2-1 line emissivity at each grid point of the simulation was
calculated from the local density and temperature according to the
analytical formulae of McKee et al.\ (\cite{mckee}), which take into
account sub-thermalization of the levels.  Emission was assumed
optically thin, which is probably a good approximation for velocities
above a few \kms.

Figure \ref{mv_iv} shows that the resulting intensity-velocity
relation for the CO J=2--1 line, $I_{\rm CO}(v)$, follows the same
slope as $m_{\rm H_2}(v)$ at low velocities, but breaks to an even
steeper relation than $m_{\rm H_2}(v)$ at velocities above $\simeq 30$
\kms.  This steepening results from the temperature dependence of the
CO(2-1) emissivity per molecule.  This dependence (assuming LTE) is
given in terms of a function $\Omega(T)$ in Cabrit \& Bertout
(\cite{c&b}), and is plotted as a function of temperature for the CO J=2-1 
line in Fig.~13 of Lada \& Fich (\cite{LF}). The line emissivity per molecule is
seen to initially increase steeply with $T$ up to a maximum at $T
\simeq$ 20~K, similar to the excitation energy of the upper level of
the transition.  For higher temperatures, however, it then {\em
decreases as $T^{-1}$}, due to the larger number of energy levels
available to the molecule (or equivalently, as an effect of the partition
function). Hence, as we go to higher velocities, not only are there
fewer molecules to emit (due to shock dissociation), but they are
hotter (because they have been through a stronger shock) and so they emit
less efficiently in the CO J=2-1 line. This produces an even steeper
slope at high velocities in $I_{\rm CO}(v)$ than in $m_{\rm H_2}(v)$.

\subsection{The $I_{\rm H_2}(v)$ relation}
\label{num_h2_relation}

It is interesting to explore the predicted intensity-velocity relation for
the H$_2$ S(1) 1--0 line, $I_{\rm H_2}(v)$, as it has been recently
observed in a few flows (Salas \& Cr\'uz-Gonz\'alez \cite{salas}) and the
line is of much higher excitation than the CO J=2-1 line. Such predictions
from jet simulations have, to the best of our knowledge, never been
presented so far. For simplicity, we computed the line emissivity assuming
LTE. Although this is not fully justified at the moderate densities of our
simulations, it is sufficient to illustrate the change of slope with
excitation of the line that we predict to occur in jet-driven outflows.

From Fig.\ \ref{mv_iv} we see that, unlike the CO 2-1 line, this 
intensity-velocity relation is {\it shallower} than
$m_{\rm H_2}(v)$ up to 30 \kms, and therefore does not show as
dramatic a break in slope at high velocities. This is due to the fact
that the the upper energy level for the H$_2$ S(1) 1--0 line lies at
7000~K; hence the emissivity per molecule is still increasing with $T$
up to temperatures of $\simeq 10^4$ K, so that higher velocity, hotter
material now emits more efficiently in the H$_2$ S(1) 1--0 line.  The 
$I_{\rm H_2}(v)$ distribution is then shallower than the 
underlying mass distribution.

\section{Comparison with observations}
\label{comp_obs}

\begin{figure}
\resizebox{\hsize}{!}{\includegraphics{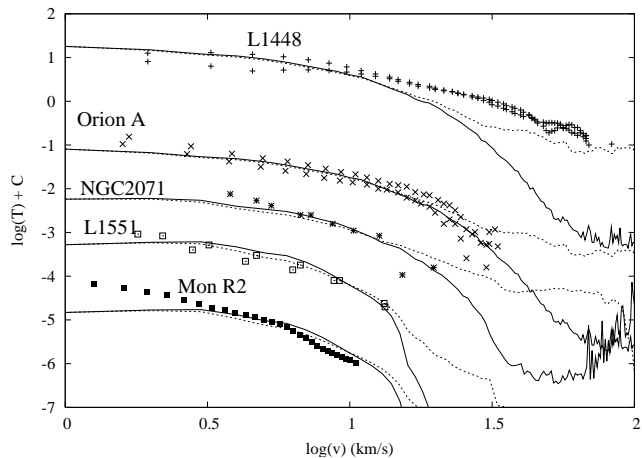}}
\caption{Plots of the observed intensity-velocity relations for L1448,
Orion A, NGC2071, L1551 and Mon R2 (`+' signs, crosses, stars, open
boxes, and filled boxes respectively).  Also shown are the simulated
$m(v)$ and $I_{\rm CO}(v)$ (dotted and solid lines, respectively, with arbitrary
vertical offsets) for comparison.  The fits to both L1448
and Orion A assume an angle of $60^\circ$ to the plane of the sky, for
NGC2071 the angle is assumed to be $30^\circ$, and for L1551 and Mon R2
the flows are assumed to be in the plane of the sky.
\label{sim_obs}}
\end{figure}

In this section, we compare our simulation results directly with
observations.  Figure \ref{sim_obs} contains plots of our results for the
$m(v)$ and $I_{\rm CO}(v)$ relations, along with observed
intensity-velocity relations in CO J=2-1 for various bipolar outflows
(taken from Bachiller \& Tafalla \cite{bataf}). A similar
comparison with the results of Salas \& Cr\'uz-Gonz\'alez (\cite{salas})
will be made in a forthcoming paper, using NLTE calculations 
of the H$_2$ S(1) 1--0 line intensity.

It is notable that the simulation results for the
$I_{\mathrm{CO}}(v)$ relation reproduce the observations remarkably well,
given that the velocity of the jet was arbitrarily chosen and the
simulation timescale is shorter than that of the physical
systems (estimated ages range from 1000 yrs in Orion to several times $10^4$ yrs
for L~1551 and Mon R2). It is also interesting to note that the predicted
$I_{\mathrm{CO}}(v)$ more closely matches the observations than the
total mass-velocity relation $m(v)$, save for L1448.

The only parameters which have been tuned in this comparison are the
vertical scale (i.e. a scaling in the ambient gas density), and the
assumed inclination to the plane of the sky.  The latter
parameter can be seen to have a sizeable effect on the break velocity.
While our adopted inclinations do not exactly match that inferred from
observations for each of the systems shown, the trend is correct in
the sense that Orion A and L1448 are the most inclined to the plane of
the sky, NGC2071 is intermediate, and L1551 and Mon R2 are believed to
be closest to the plane of the sky.

In summary, it seems clear, on the basis of these simulations, that a
jet-driven bowshock can reproduce the observed $I_{\mathrm{CO}}(v)$
remarkably well, both in terms of the slopes at low and high
velocities and in terms of the range of break velocities where this
change of slope occurs, without the need to invoke two distinct entrainment
mechanisms as is done in Zhang \& Zheng (\cite{zz97}).

The simulation results do tend to be slightly too flat at the lowest
velocities, especially in the oldest flows (Mon R2 in particular). This 
could be due to the short timescales in our simulations, as we now show.

\section{The effect of age and time-variability}
\label{effects}

\subsection{Effect of time-variability of the jet velocity}

\begin{figure}
\resizebox{\hsize}{!}{\includegraphics{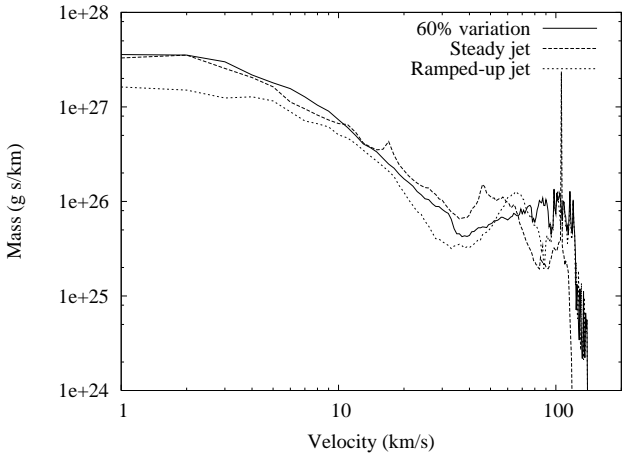}}
\caption{Plots of the mass-velocity distribution of swept-up material
for three simulations with differing jet velocity variability at
$t=400$ yrs, assuming an angle of $30^\circ$ to the plane of the sky.
Note the similarity between all three relations.
\label{mv_all}}
\end{figure}

It is important to verify that our results are not critically
dependent upon the adopted jet velocity behaviour. Figure \ref{mv_all}
compares the $m(v)$ relation for the pulsed jet with the swept-up mass-velocity
relation for the other two simulations (steady jet and ramped-up jet),
at the same time and same inclination. It is clear that the
relations are very similar.  This is also the case for the $m_{\rm
H_2}(v)$, $I_{\rm CO}(v)$ and $I_{\rm H_2}(v)$ relations (not shown).
This strongly suggests that variability of the jet velocity (at least
on time-scales less than about 50 yrs) will not greatly affect the
resulting mass-velocity or intensity-velocity relations.

\subsection{Effect of age}
\label{age_effect}
\begin{figure}
\resizebox{\hsize}{!}{\includegraphics{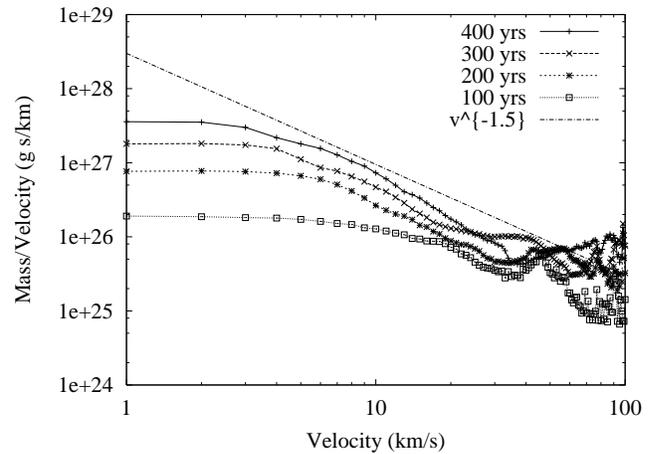}}
\caption{Plots of the mass-velocity relation for all swept-up material in
the pulsed jet (simulation (2)) at $t=100$, 200, 300 and 400 yrs, assuming an angle of 
$30^\circ$ to the plane of the sky.  \label{mv_temporal}}
\end{figure}

It is also important to verify that our results do not depend critically on the
simulation timescale, since observed flows cover a large range of ages.
Figure \ref{mv_temporal} contains plots of the mass-velocity relation
for the pulsed jet simulation (2) at different times from $t=100$--400 yrs.
It can be seen that the part of the $m(v)$ distribution that follows a
power-law $\propto v^{-1.5}$ extends to progressively lower velocities
as time goes, while the flat portion of $m(v)$ at very low velocity
shrinks. This behaviour can be understood as follows: the power-law
regime corresponds to the part of the bow structure presently on the
grid, while the flat part of $m(v)$ comes from the truncation of the
bow at the left border of the computational grid; as the bow head
advances, more of the bow structure appears on the grid, so that the
part affected by truncation is at a lower speed (since velocity
decreases away from the bow head). Hence, we expect
that for longer timescales, the flat part of $m(v)$ and $I_{\rm
CO}(v)$ will fall to very low velocities, while the $v^{-1.5}$
power-law should extend over the whole observable velocity range, in
better agreement with observations of old flows such as Mon R2.
Simulations over very long timescales are under way to study this
hypothesis.

\section{Analytic model}
\label{analytic_model}
\begin{figure}
\resizebox{\hsize}{!}{\includegraphics{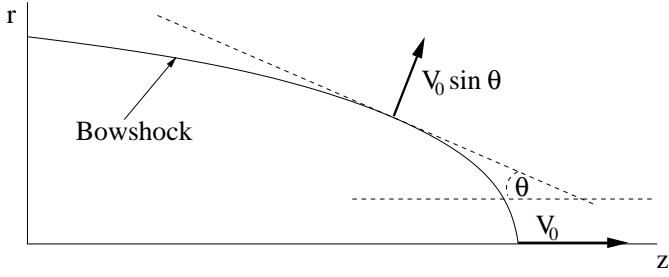}}
\caption{Schematic diagram of the set-up used in the analytic
calculations of the mass-velocity relation.
\label{analytic_setup}}
\end{figure}

In this section we explore in more depth how a jet-driven bowshock can
produce the power-law slope $\mu \simeq 1.5$ of the mass-velocity
relation discussed in Sect.\ \ref{num_totalmv}.  Figure
\ref{analytic_setup} contains a schematic diagram of the set-up used
in this discussion.

Consider an idealised bowshock, the shape of which is given by the relation
$z \propto r^s$, moving in the positive $z$ direction with constant
velocity $v_0$ into an ambient medium of constant density $\rho_{\rm a}$.
We assume that there is no mixing in the bow, so that the magnitude of the
post-shock velocity in the observer's frame is $v_0 \sin \theta$, with
$\tan\theta = dr/dz$ (see Fig.\ \ref{analytic_setup}). An observer situated
along the $z$-axis will then see a line-of-sight velocity of
\begin{equation}
v=v_0\sin^2\theta
\end{equation}
Far from the apex, where $\theta$ is small, $v \propto \tan^2
\theta \propto (r^2)^{1-s}$. 
The rate of increase of mass at projected velocity $v$ is then
\begin{equation}
\frac{d}{dt}\left(m(v) dv\right)  = \rho_\mathrm{a} v_0 2 \pi r(v) dr 
 \propto \left( v \right)^{{s}/({1-s})} dv.
\end{equation}
Hence the mass-velocity relation in this case is a power-law 
\begin{equation}
m(v) \propto v^{-\mu} \quad {\rm with}\quad \mu = \frac{s}{s-1}.
\label{mu_def}
\end{equation}

In order to verify that the no-mixing model can indeed explain the $m(v)$
relations found in our simulations, we need to determine the
appropriate value of $s$ for the simulated bows.  For this, we examine the
rate at which ambient mass is swept up by the bowshock,
integrated over all velocities. For an adopted ideal bow shape 
$z \propto r^s$, this rate is
\begin{equation}
\frac{dm}{dt} (t) = \pi r_0^2 \rho_{\rm a} v_0
\end{equation}
where $r_0 \propto z^{1/s}(t)$ is the maximum bowshock radius and 
$z(t)=v_0 t$ is 
the length of the bowshock, both at time $t$.  Hence, the total swept-up
mass as a function of time is predicted to vary as
\begin{equation}
m(t) \propto t^\alpha \quad {\rm with}\quad \alpha = {1+2/s}.
\end{equation}

\begin{figure}
\resizebox{\hsize}{!}{\includegraphics{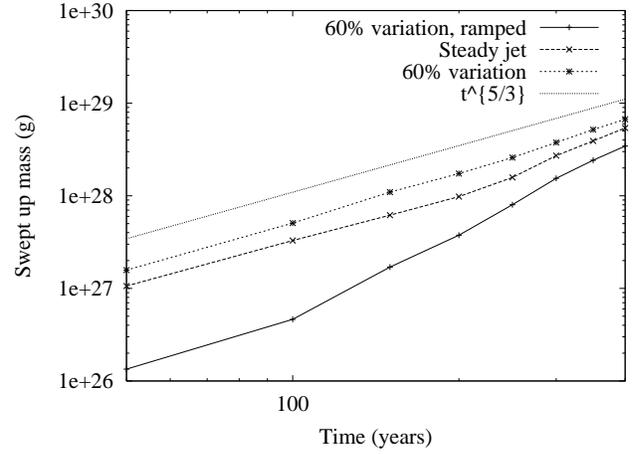}}
\caption{Plots of the total swept-up mass against time for each of the
simulations.  \label{mass_time}}
\end{figure}

In Fig.\ \ref{mass_time} we plot the total swept-up mass $m(t)$ as a
function of time for each of the simulations. Indeed, we find that it
closely follows a power-law, with a slope $\alpha=1.7$--1.8. Note that
the apparent deviation of simulation (3) from this law at early times
is due to the initial ramping up of the jet velocity $v_0$.  Since the
bow model predicts $\alpha = 1+2/s$, we infer that the appropriate
value of $s$ for our simulated bowshocks is $s=2.5$--$2.9$.  If we now
substitute this range of values of $s$ into equation \ref{mu_def}, we
find that our simple bow model with no mixing predicts an exponent of the
mass-velocity relation in the range $\mu = 1.5$--$1.7$.
This is remarkably close to what we find in the
simulations (Sect.\ \ref{num_totalmv}).
Note that the same bow model, but assuming instantaneous mixing of
post-shock swept-up gas, predicts a much steeper slope $\mu = 2 + s/2 = 
3.5$ (Smith et al.\ \cite{smith}).

\begin{figure}
\resizebox{\hsize}{!}{\includegraphics{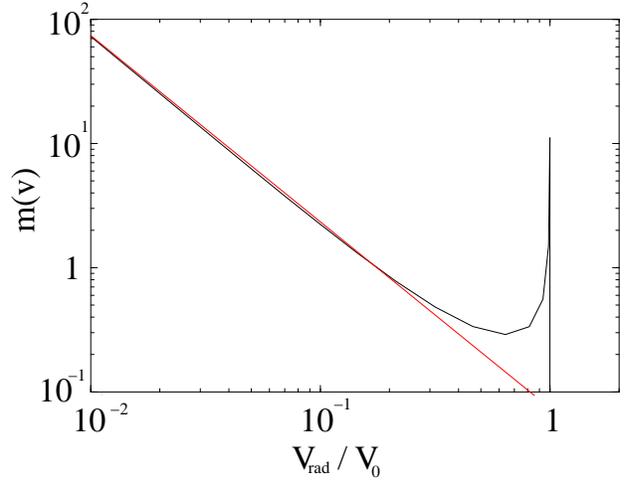}}
\caption{Plot of the total swept-up mass as a function of velocity predicted 
for an ideal bowshock shape described by $z \propto r^3$ viewed pole-on,
with no post-shock mixing. The straight line is the power law of exponent
-1.5 predicted for the bow wings (see text). \label{analytic_mv}}
\end{figure}

It is important to keep in mind that the power-law slope is predicted
to hold only for small values of $\theta$, i.e.\ sufficiently far from
the head of the bowshock. In Figure \ref{analytic_mv} we plot the
mass-velocity relation for a theoretical bow with $s = 3$, calculated
numerically without the assumption of small $\theta$. The power-law of
slope $\mu = s/(s-1)= 1.5$ is seen to give a very good approximation
to the actual $m(v)$ up to about 20\% of the jet velocity.  At higher
velocities, the mass-velocity relation flattens out and then increases
with velocity. This behaviour is clearly also seen in our simulated
$m(v)$ (see Fig.\ \ref{mv_iv}), although we do not find as sharp a
peak at $v_0$, due to the fact that the head of the bowshock is rather
irregular.

\section{Conclusions}
\label{conclusions}

We find that the mass-velocity relation in simulations of jet-driven
molecular outflows is a shallow power-law of exponent $\simeq$ -1.5, with
no break to a steeper slope at higher velocity, unlike previous analytical
predictions by Zhang \& Zheng (\cite{zz97}). This $m(v)$ relation can be
well explained by a bowshock model with no post-shock mixing. It does not
appear to be consistent with the full mixing model proposed by Smith et al.
(\cite{smith}).

We also find that the resulting intensity-velocity relation for the CO J=2-1
line compares remarkably well with observations of molecular outflows. In
particular, it {\em does} have a break in slope around 20-30 km
s$^{-1}$. The break is found to result from molecular dissociation near the
bow apex, and from the $1/T$-dependence of emission at temperatures
exceeding the energy of the upper level of the line. Because of this
dependence on $T$, a jet-driven model predicts a shallower slope at
high velocity in higher excitation lines (e.g.\ H$_2$ S(1)(1-0) and high-J
CO lines), which could be tested by ongoing studies.  Another
implication of our results is that, in jet-driven outflows, the CO J=2-1 
intensity profile reflects the slope of the underlying mass-velocity 
distribution only at velocities $\le $ 20 \kms, and that higher temperature 
tracers are required to probe the mass distribution at higher speed.

The only weak point of a bowshock model with no mixing is that emitting gas
expands perpendicular to the bow surface, and thus there is some difficulty in
explaining the apparent ``forward-directed'' motion of molecular outflows
noted by Lada \& Fich (\cite{LF}). Quantitatively, our simulations predict
comparable amounts of redshifted and blueshifted gas up to velocities of
3-4 \kms. Full mixing was invoked by Smith et al. (\cite{smith}) to 
alleviate this problem, but we have shown here that it would predict a very 
steep $m(v)$ of slope $\simeq -3.5$ and an even steeper CO intensity-velocity 
relation (due to dissociation and temperature effects), inconsistent with
observations. We note that another situation in which motion will be
more forward-directed is when the bowshock propagates into already moving 
material. This is an interesting issue which clearly requires further work.

\begin{acknowledgements}
We would like to thank Rafael Bachiller for kindly providing us with the 
observational data in Figure \ref{sim_obs} and for stimulating discussions.
We are indebted to Alex Raga for his contribution to the code calculating 
CO emissivities, and for helpful suggestions about this work.
T.P. Downes acknowledges support as a visiting astronomer at the
Observatoire de Paris.
\end{acknowledgements}

\end{document}